# A Facile Method for Precise Layer Number Identification of Two-Dimensional Materials through Optical Images

Jiayu Lei[1], Jiafan Qu[1], Peng Wang[1], Hu Jiang[1], Hongyan Shi[1,2], Xiudong Sun[1,2], and Bo Gao[1,2,a]

[1] Institute of Modern Optics, Department of Physics, Key Laboratory of Micro-Nano Optoelectronic Information System, Ministry of Industry and Information Technology, Key Laboratory of Micro-Optics and Photonic Technology of Heilongjiang Province, Harbin Institute of Technology, Harbin 150001, China

[2] Collaborative Innovation Center of Extreme Optics, Shanxi University, Taiyuan 030006, China

[a] Author to whom correspondence should be addressed. Electronic mail: gaobo@hit.edu.cn

## Abstract

Optical microscopy is believed to be an efficient method for identifying layer number of two-dimensional 2D materials. However, since illuminants, cameras and their parameters are different from lab to lab, it is impossible to identify layer numbers just by comparing a given optical image with standard or calculated images under standard conditions. Here we reported an image reconstruction method, converting raw optical images acquired by arbitrary illuminants and cameras into reconstructed images at specified illuminant and specified camera. After image reconstruction, the color differences of each layer number roughly equaled those calculated under specified condition. By comparing the color differences in reconstructed image with those calculated under specified condition, the layer numbers of 2D materials in our lab and published papers, including $MoS_2$, $WS_2$ and $WSe_2$, were ambiguously identified. This study makes optical

---

[a] **Author to whom correspondence should be addressed. Electronic mail: gaobo@hit.edu.cn**

microscopy a precise method for identifying layer numbers of 2D materials on known substrate.

## Introduction

Since the discovery of monolayer graphene in 2004, various two-dimensional (2D) materials, such as hexagonal boron nitride (hBN), transition metal dichalcogenides (TMDs) and black phosphorus (BP), have attracted a remarkable interest due to potential applications in thin-film electronics and optoelectronics. Generally, physical properties of 2D materials are closely correlated with their electronic structures, which are greatly influenced by their layer numbers [1-4]. Experimentally, atomic force microscopy (AFM) [5, 6] and Raman spectroscopy [7, 8] are the most common methods for layer number identification. Despite of high reliability, these methods require time-consuming procedures and operations for large scale 2D materials. A lot of efforts have been made to develop new methods in both high reliability and large scale. Therein, optical microscopy is believed to be a suitable method. Based on the interference theory, the optical contrast between substrate (usually a $SiO_2$/Si substrate with known oxide thickness) and 2D material is considered to be a function of layer numbers, which can be utilized to identify layer numbers [9-12]. The optical contrast distribution is actually a grey-scale image with a range from zero to one, while an optical RGB image retains additional degrees of spectral information. To achieve more precise results, each of the RGB channels of an image is supposed to take into account. So far, color differences in CIELab color space [13], sRGB color space [14-17] and CIE XYZ color space [18] have been calculated to quantitatively analyze layer numbers of 2D materials.

The optical image of 2D materials is mainly influenced by the reflectance of the object calculated with corresponding refractive index [19], the spectrum of the illuminant, the camera and its setups (such as white balance and exposure). However, as illuminants and cameras are different from lab to lab, it is impossible to determine layer numbers by comparing with published images. Therefore, although theoretically feasible, optical microscopy has never been a by-itself layer number identification method for 2D materials.

In this study, we converted raw optical images acquired by arbitrary illuminants and cameras to new images at specified illuminant and specified camera by image reconstruction. After image reconstruction, the color differences roughly equaled those calculated under specified condition for specific 2D materials on known substrate, regardless of illuminant and camera. This method, bypassing the illuminants and cameras, provides a precise tool to identify layer numbers of thin 2D materials.

## Color difference and image reconstruction

The color of an image is usually studied in color space, e.g. CIE (International Commission on Illumination) XYZ color space, where a certain color is described in three-dimensional coordinate. By CIE color-matching equation, the color of an object in CIE XYZ color space is given as follows [20]:

$$\begin{aligned}
X &= k \int_{380 \text{ nm}}^{780 \text{ nm}} R(\lambda)I(\lambda)\,\bar{x}(\lambda)d\lambda \\
Y &= k \int_{380 \text{ nm}}^{780 \text{ nm}} R(\lambda)I(\lambda)\,\bar{y}(\lambda)d\lambda \\
Z &= k \int_{380 \text{ nm}}^{780 \text{ nm}} R(\lambda)I(\lambda)\,\bar{z}(\lambda)d\lambda \\
k &= \frac{100}{\int_{380 \text{ nm}}^{780 \text{ nm}} I(\lambda)\bar{y}(\lambda)d\lambda}
\end{aligned} \quad (1)$$

where $X$, $Y$, and $Z$ are tristimulus values which describe the color of an object (such as 2D materials nanosheets) in CIE XYZ color space. $k$ is the normalization coefficient. $R(\lambda)$ is the spectral reflectance of the object, which is related to the film thickness due to thin-film interference [9, 21]. $I(\lambda)$ is the spectral power distribution (SPD) of the illuminant. $\bar{x}(\lambda)$, $\bar{y}(\lambda)$ and $\bar{z}(\lambda)$ are the color matching functions of the camera or observer.

Then, the color difference between substrate and 2D materials can be expressed as $\Delta E_{XYZ} = \sqrt{(X_{2D} - X_s)^2 + (Y_{2D} - Y_s)^2 + (Z_{2D} - Z_s)^2}$, where $(X_{2D}, Y_{2D}, Z_{2D})$ and $(X_s, Y_s, Z_s)$ are color coordinates of 2D materials and substrates, respectively. Since CIELab color space with CIE76 definition is designed to be more perceptually uniform to human color vision than CIE XYZ color space, color difference between substrate and 2D materials is usually defined as

$$\Delta E_{ab}^* = \sqrt{(L_{2D} - L_s)^2 + (a_{2D} - a_s)^2 + (b_{2D} - b_s)^2} \quad (2)$$

where $(L_{2D}, a_{2D}, b_{2D})$ and $(L_s, a_s, b_s)$ are color coordinates of 2D materials and substrates in CIELab color space, respectively. Thus, if SPD, color mathcing functions and oxide thickness are known, we can directly obtain the layer number from $\Delta E_{ab}^*$.

In practice, parameters such as SPD and color mathcing functions in an optical microscopy are usually undefined. So we manage to convert raw image under arbitrary condition to new image under a given condition. To achieve larger color differences, we specify CIE Standard Illuminant A [22] as illuminant and CIE standard observer [22] as camera, which is defined as specified condition in this work (See Figure S1 for calculated color image under specified condition and other conditions in Supplementary Material).

For a 2D nanosheet on $SiO_2$/Si substrate with known oxide thickness, the reflectance spectrum and hence the color of the substrate are fixed under specified condition. So the color of a substrate can be referred to reconstruct the raw image under arbitrary conditions.

The color of each pixel in an image can be expressed as a three elements' vector. For simplicity, we reformulated Eq. 1 in terms of matrix product:

$$\mathbf{t} = \mathbf{ALr} \qquad (3)$$

where $\mathbf{t} = [X\ Y\ Z]^T$ contains tristimulus values, the columns of the $3 \times N$ matrix $\mathbf{A}$ include the CIE XYZ color matching functions; $\mathbf{L}$ is an diagonal matrix whose diagonal elements are the spectral power distribution; $N$ elements' vector $\mathbf{r}$ is the spectral reflectance of an object. Therefore, XYZ tristimulus values under specified condition are

$$\mathbf{t}_{spe} = \mathbf{A}_{spe}\mathbf{L}_{spe}\mathbf{r} \qquad (4)$$

For raw image with arbitrary illuminant and color matching functions, tristimulus values are

$$\mathbf{t}_{raw} = \mathbf{A}_{raw}\mathbf{L}_{raw}\mathbf{r} \qquad (5)$$

As mentioned above, the color of the substrate under specified condition is regarded as reference. Then, by using differential XYZ tristimulus values of substrate under specified condition and in raw image, we can reconstruct the whole raw image and obtain

XYZ tristimulus values of reconstructed image by Eq. 6 [23]:

$$\mathbf{t}_{rec} = \mathbf{t}_{raw} + \alpha J(\mathbf{t}_{spesub} - \mathbf{t}_{rawsub}) \tag{6}$$

where $\mathbf{t}_{spesub}$ and $\mathbf{t}_{rawsub}$ represent XYZ tristimulus values of substrate under specified condition and raw image; $\mathbf{t}_{raw}$ is the tristimulus values of 2D materials in raw image; $\mathbf{t}_{rec}$ is the tristimulus values of reconstructed image; $\alpha J$ is a nonlinear transformation. In this work, $\alpha J = 1$ is used, as we do not know the exact relationship between color and layer number in a raw image.

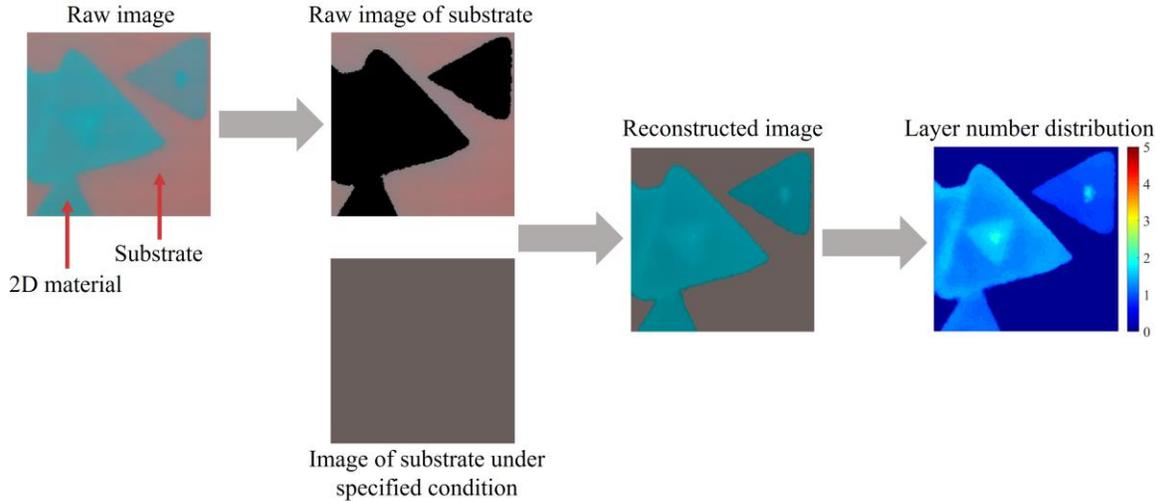

**Figure 1**. Procedures of image reconstruction and layer number distribution calculation.

Based on the method above, the optical image reconstruction and layer number distribution calculation procedure is shown in Figure 1. Firstly, we choose the substrate domain in a raw image and then calculate the average sRGB values of the substrate, which are subsequently converted to XYZ tristimulus values. After calculating XYZ tristimulus values of substrate under specified condition, we get differential XYZ tristimulus values of substrate under specified condition and in raw image. Then, the differential XYZ tristimulus values are applied to each pixels of the raw image based on Eq. 6, so that we can obtain the XYZ tristimulus values of the reconstructed image. In the end, the obtained tristimulus values of the whole image are converted back to sRGB values for display and CIELab values for layer number distribution calculation. Based on Eqs. 1-2 to calculate the color difference relationship between substrate and nanosheets under specified condition, the layer number of each pixel, and in a larger scale, each domain, can be finally identified.

**Layer number identification from reconstructed calculated image**

To demonstrate the feasibility, we calculated three optical images of $MoS_2$ nanosheets consisting of 1-5 layers on Si substrate with 300 nm $SiO_2$ layer ($SiO_2$/Si) under three selected conditions and then reconstructed the three images. In Figure 2a we used Illuminant D65 [22] as illuminant. In Figure 2b we used MT9V032 [24] CMOS as camera. In Figure 2c we used Gray World algorithm [25] for white balance calibration. It can be seen that there are different contrasts between substrate and different layer numbers in all three images. Figure 2d shows the color difference versus layer number extracted from the three images (color difference values are shown in Table SI in Supplementary Material) and the image under specified condition (See Figure S1a in Supplementary Material). It can be seen that the color differences are increasing with layer numbers, which indicates that layer number can be identified by color difference. However, it is noted that there are remarkable differences between color differences of each layer number in the four images.

Therefore, it is highly possible to inaccurately identify layer numbers by comparing color differences extracted from optical images acquired under different conditions.

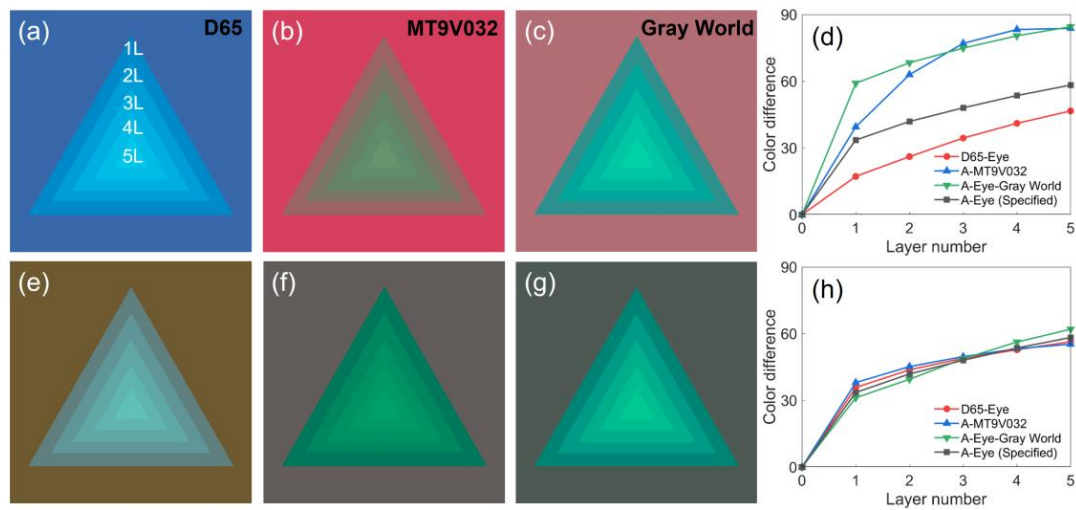

**Figure 2**. Calculated optical images of MoS$_2$ nanosheets consisting of 1-5 layers on SiO$_2$/Si substrate, with illuminant D65 and CIE standard observer (a) before (e) after image reconstruction; with illuminant A and EO-0413C Camera with uEye MT9V032 CMOS (Edmund Optics) (b) before (f) after image reconstruction; with illuminant A and CIE standard observer, and Gray World AWB algorithm calibration (c) before (g) after image reconstruction. Color difference between MoS$_2$ nanosheets and SiO$_2$/Si substrate versus layer number (d) before (h) after image reconstruction.

Figure 2e-2f show the reconstructed images from Figure 2a-2c, respectively. Figure 2h shows the color differences versus layer numbers extracted from the three reconstructed images and the image under specified condition (color difference values are shown in Table SI in Supplementary Material). It can be seen that the color differences extracted from the reconstructed images are roughly equal to that extracted from the image under specified

condition. More examples under different conditions are shown in Figure S2 (See Supplementary Material). Therefore, after image reconstruction, we could accurately identify layer numbers by comparing color differences extracted from reconstructed images.

**Layer number identification from reconstructed experimental image**

To testify the universality of the image reconstruction method, we reconstructed some optical images of 2D materials in our lab and in published papers. Figure 3a and 3b show the raw and reconstructed optical images of our $MoS_2$ nanosheets sample on Si substrate with 300 nm thick $SiO_2$ layer prepared by CVD method. Optical images were obtained by a EO-0413C camera (Edmund Optics) with uEye MT9V032 CMOS attached to an Olympus BX51 optical microscope under Olympus U-LH100-3 halogen illuminant. Raman measurement shows that there are monolayer, bilayer and trilayer $MoS_2$ nanoseets (See Figure S3 in Supplementary Material), as indicated in Figure 3a.

The purple curves and black curves in Figure 3c show the color difference versus layer number extracted from Figure 3a and Figure S1a, respectively (color difference values are shown in Table SII in Supplementary Material). It can be seen that color differences are increasing with layer numbers, and there are remarkable differences between color differences of each layer number under experimental condition and specified condition. e.g., color differences of monolayer $MoS_2$ in Figure 3a and Figure S1a are 55.4 and 33.6. After reconstruction, as shown in Figure 3d, all the color differences of each layer number in reconstructed optical image roughly equal those under specified condition. e.g., color differences of monolayer MoS$_2$ is 36.6, close to 33.6. We also reconstructed three published optical images of MoS$_2$ on Si substrate with 300 nm SiO$_2$ layer (see reconstructed images in Figure S4 in Supplementary Material), and plotted the color difference versus layer number for raw image and reconstructed image in Figure 3c and 3d, respectively. For raw images, the color differences of each layer number have a broad range. e.g., for monolayer, color differences range from 16.3 to 43.1. After image reconstruction, all the color differences of each layer number are roughly equal to those under specified condition. e.g., for monolayer, color differences range from 32.5 to 34.0, close to 33.6. Therefore, we can identify layer numbers of MoS$_2$ nanosheets just from reconstructed optical images, even though the illuminants and cameras are unknown.

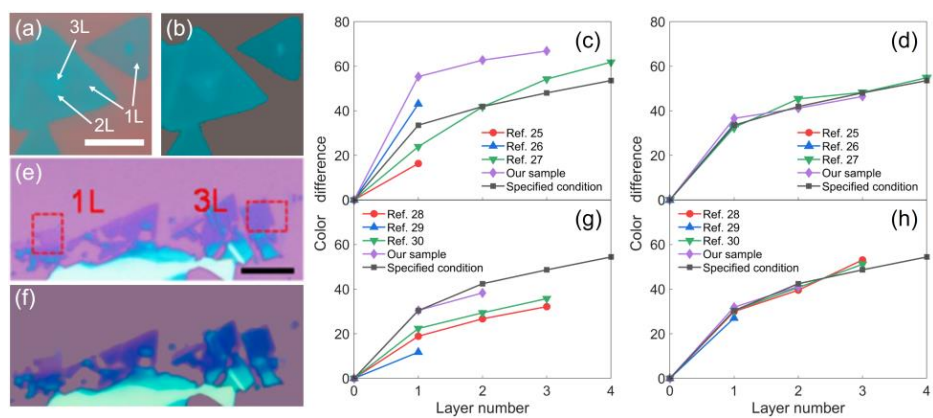

**Figure 3**. Optical images of our MoS$_2$ sample (a) before (b) after image reconstruction, and color difference between MoS$_2$ nanosheets [26-28] and SiO$_2$/Si substrate versus layer number (c) before (d) after image reconstruction. Scale bar: 25 μm. Optical images of WS$_2$ nanosheets [29] (e) before (f) after image reconstruction, and color difference between WS$_2$ nanosheets [29-31] and SiO$_2$/Si substrate versus layer number (g) before (h) after image reconstruction. Scale bar: 5 μm.

We also applied the image reconstruction method to other 2D materials. Figure 3a and 3b show the raw and reconstructed optical images of $MoS_2$ nanosheets on Si substrate with 300 nm thick $SiO_2$ layer [29]. Figure 3g and 3h show the color differences in raw image and reconstructed image versus layer number for $WS_2$ nanosheets in our lab and in published papers [29-31] (See raw and reconstructed images in Figure S4 and color difference values in Table SIII in Supplementary Material). For raw images, the color differences of each layer number of $WS_2$ have broad ranges due to different illuminants and cameras. After image reconstruction, all the color differences of each layer number are very close to those in specified condition. Similar results were also found in $WSe_2$ nanosheets [10, 32] (See Figure S5 in Supplementary Material). Therefore, optical microscopy could be a precise by-oneself method for identifying layer numbers of various 2D materials through image reconstruction.

## Conclusion

In this work, we develop an image reconstruction method to identify layer numbers of 2D materials through image reconstruction. With our method, raw optical images acquired by arbitrary illuminants and cameras were reconstructed into new images at specified illuminant and specified camera. After image reconstruction, the color differences of each layer number were getting close to those calculated under specified condition. By comparing the color differences in reconstructed image with those calculated under specified condition, the layer numbers of 2D materials in our lab and published papers, including $MoS_2$, $WS_2$ and $WSe_2$, were ambiguously identified. This study makes optical microscope a precise method for identifying layer numbers of 2D materials on known substrate.

## Supplementary material

See supplementary material for: Calculated optical images under different conditions; Color difference between $MoS_2$ nanosheets and $SiO_2$/Si substrate versus layer number by different illuminants, color matching functions and AWB algorithms before and after image reconstruction; Raman spectra of our $MoS_2$ and $WS_2$ samples; Raw images from published papers, and corresponding reconstructed images and layer number images; optical images and color difference of $WSe_2$; Tables of color difference values. Our codes are also included in supplementary material.

## Acknowledgements

We appreciate Yuege Xie from UT-Austin for helpful discussions. This study is financially supported by the National Natural Science Foundation of China (No. 21473046) and the New Faculty Start-up Funds from Harbin Institute of Technology.